\documentstyle[12pt,twoside,fleqn,espcrc1]{article}

\newcommand{\AmS}{{\protect\the\textfont2
  A\kern-.1667em\lower.5ex\hbox{M}\kern-.125emS}}

\title{ Near threshold $\Lambda$ and $\Sigma$ production in 
$pp$ collisions}

\author{A.M. Gasparian$^{a,b}$, J. Haidenbauer$^a$, C. Hanhart$^c$,
L. Kondratyuk$^b$, and J. Speth$^a$
\address {$^a$Institut f\"{u}r Kernphysik, Forschungszentrum J\"{u}lich
GmbH, D--52425 J\"{u}lich, Germany \\
\vskip 0.5cm 
$^b$Institute for Theoretical and Experimental Physics,
117258, B.Cheremushkinskaya 25, Moscow, Russia \\
\vskip 0.5cm 
$^c$Nuclear Theory Group and INT, Dept. of Physics,
University of Washington, Seattle, WA 98195-1560, USA 
}}

\begin{document}
\maketitle

\begin{abstract}
A model calculation for the reactions $pp\rightarrow p\Lambda K^+$ and 
$pp\rightarrow N\Sigma K$ near their thresholds is presented. 
It is argued that the experimentally observed strong suppression of
$\Sigma^0$ production compared to $\Lambda$ production at the same
excess energy could be due to a destructive interference between 
the $\pi$ and $K$ exchange contributions in the reaction  
$pp\rightarrow p\Sigma^0 K^+$. Predictions for $pp\rightarrow p\Sigma^+ K^0$ 
and $pp\rightarrow n\Sigma^+ K^+$ are given. 
\end{abstract}

\vskip 0.5cm 

In a recent measurement of the reactions $pp\rightarrow
p\Lambda K^+$ and $pp\rightarrow p\Sigma^0K^+$ 
near their thresholds it was found that the cross section for 
$\Sigma^0$ production is about a factor of 30 smaller than the one for
$\Lambda$ production \cite{Sew}.
We want to report on an exploratory investigation of the
origin of this strong suppression of the near-threshold 
$\Sigma^0$ production \cite{Gas}.
In particular we want to examine a possible explanation that was suggested 
in Ref. \cite{Sew}, namely effects from the strong $\Sigma N$ final 
state interaction (FSI) leading to a $\Sigma N \rightarrow \Lambda N$ 
conversion. 
We treat the associated strangeness production in the standard
distorted wave Born approximation. 
We assume that the strangeness production process is governed by the
$\pi$- and $K$ exchange mechanisms.
In order to have a solid basis for our study of possible conversion 
effects we employ a microscopic $YN$ interaction model developed
by the J\"ulich group (specifically model A of Ref. \cite{Holz}).
This model is derived in the meson-exchange picture and takes into
account the coupling between the $\Lambda N$ and $\Sigma N$ channels. 
 
The vertex parameters (coupling constants, form factors) appearing at 
the $\pi NN$ and $K NY$ vertices in the production diagrams 
are taken over from the J\"ulich $YN$ interaction.
The elementary amplitudes $T_{KN}$ and $T_{\pi N \rightarrow KY}$
are taken from corresponding microscopic models \cite{Hof1,Hof2}  
that were developed by our group. However, for simplicity reasons 
we use the scattering length and on-shell threshold amplitudes, 
respectively, instead of the full (off-shell) $KN$ and 
$\pi N \rightarrow KY$ t-matrices. The off-shell extrapolation 
of the amplitudes is done by multiplying those quantities 
with the same form factor that is used at the vertex where the
exchanged meson is emitted. Only s-waves are considered. 

We do not take into account the initial state interaction (ISI)
between the protons. Therefore we expect an overestimation of the cross 
sections by a factor of around 3 in our calculation \cite{Bat}. 
But since the thresholds for
the $\Lambda$ and $\Sigma^0$ production are relatively close together
and the energy dependence of the $NN$ interaction is relatively
weak in this energy region the ISI effects should be very
similar for the two strangeness production channels and therefore
should roughly drop out when ratios of the cross sections are taken.

The cross section ratio for $K$ exchange alone and 
based on the Born diagram is 16, cf. Table 1. 
Including the $YN$ FSI, i.e. possible conversion effects 
$\Sigma N \rightarrow \Lambda N$, leads to a strong enhancement of 
the cross section in the $\Lambda$ channel but only to a
moderate enhancement in the $\Sigma^0$ channel. As a consequence, the 
resulting cross section ratio becomes significantly larger than the 
value obtained from the Born term and, in fact, exceeds the 
experimental value. In case of pion exchange the Born diagram yields a 
cross section ratio of 0.9. Adding the FSI increases the cross section 
ratio somewhat, but it remains far below the experiment.

Thus, it's clear that, in principle, $K$ exchange alone
could explain the cross section ratio - especially after 
inclusion of FSI effects. However, we also see from 
Table 1 that $\pi$ exchange is possibly the 
dominant production mechanism for the $\Sigma^0$ channel
and therefore it cannot be neglected. Indeed, 
the two production mechanisms play quite different roles in the 
two reactions under consideration. $K$ exchange yields by far the 
dominant contribution for $pp\rightarrow p\Lambda K^+$. 
The influence from $\pi$ exchange is very small. 
In case of the reaction $pp\rightarrow p\Sigma^0K^+$,
however, $\pi$- and $K$ exchange give rise to contributions
of comparable magnitude. This feature becomes very important when
we now add the two contributions coherently and consider different 
choices for the relative sign between the $\pi$ and $K$ exchange
amplitudes. In one case (indicated by ``$K+\pi$'' in Table 1)
the $\pi$ and $K$ exchange contributions
add up constructively for $pp\rightarrow p\Sigma^0K^+$ and
the resulting total cross section is significantly larger than
the individual results. For the other choice (indicated by 
``$K-\pi$'') we get a destructive interference between the amplitudes
yielding a total cross section that is much smaller.
Consequently, in the latter case the cross section ratio is
much larger and, as a matter of facts, in rough agreement with the 
experiment (cf. Table 1) - suggesting a destructive interference
between the $\pi$ and $K$ exchange contributions as a possible
explanation for the observed suppression of near-threshold
$\Sigma^0$ production. 
  
\medskip

\noindent Table 1: Total cross section of the reactions 
$pp\rightarrow p\Lambda K^+$ ($\sigma_\Lambda$)
and $pp\rightarrow p\Sigma^0K^+$ ($\sigma_{\Sigma^0}$) at
the excess energies $Q$ = 13.2 MeV ($\sigma_\Lambda$)
and $Q$ = 13.0 MeV ($\sigma_{\Sigma^0}$). 
\par\bigskip
\hfill\vbox{{\halign
{\quad#\hfil&\quad\hfil#\hfil&
\quad\hfil#\hfil&\quad\hfil#\hfil\cr
\noalign{\smallskip\hrule\smallskip}
diagrams  & $\sigma_{\Lambda}\ [nb]$ 
& $\sigma_{\Sigma^0}\  [nb]$ 
& $\frac{\sigma_{\Lambda}}{\sigma_{\Sigma^0}}$ \cr
\noalign{\smallskip\hrule\smallskip}
K (Born) & 739 & 46 & 16 \cr
K (FSI) &2426 & 57 & 43 \cr
\noalign{\smallskip\hrule\smallskip}
$\pi$ (Born) & 71 & 77 & 0.9 \cr
$\pi$ (FSI) & 113 & 105 & 1.1 \cr
\noalign{\smallskip\hrule\smallskip}
``$K + \pi$'' (FSI) & 2471 & 251 & 9.9 \cr
``$K - \pi$'' (FSI) & 2607 & 73 & 36 \cr
\noalign{\smallskip\hrule\smallskip}
exp. [1] & 505$\pm$ 33 & 20.1$\pm$3.0 & 25$\pm$6 \cr
\noalign{\smallskip\hrule\smallskip}
}}}\hfill\break\par
\par\smallskip

It is now interesting to look also at corresponding results for other 
$\Sigma$ production channels. E.g., for the reaction 
$pp\rightarrow n\Sigma^+ K^+$ the predicted cross sections at the
excess energy of 13 MeV are 86 (``$K+\pi$'') and
229 $nb$ (``$K-\pi$''), respectively. Thus, the interference
pattern is just the opposite as for $pp\rightarrow p\Sigma^0K^+$,
cf. Table 1.
For the ``$K-\pi$'' case favoured by the experimental
$\sigma_\Lambda/\sigma_{\Sigma^0}$ ratio our calculation yields a cross
section for $pp\rightarrow n\Sigma^+K^+$ that is about 3 times
larger then the one for $pp\rightarrow p\Sigma^0K^+$. Such a ratio
is in fair agreement with data and model calculations at higher
energies, see, e.g. Ref.~\cite{Lag}. The other choice, ``$K+\pi$'',
leads to a $\sigma_{pp\rightarrow n\Sigma^+K^+}$ that is a factor of about 3 
smaller than $\sigma_{pp\rightarrow p\Sigma^0K^+}$ - 
a result which is rather difficult to reconcile
with the present knowledge about these reactions at higher energies.
These features are displayed graphically in Fig. 1 as a function of the
excess energy. Note that all curves are multiplied by a common reduction
factor of 0.3 \cite{Gas} to compensate for ISI effects \cite{Bat}. 

For the reaction $pp\rightarrow p\Sigma^+ K^0$ the predicted cross 
sections at the excess energy of 13 MeV are 725 (``$K+\pi$'') and
423 $nb$ (``$K-\pi$''), respectively. Thus, in this case the interference
pattern is the same as for $pp\rightarrow p\Sigma^0K^+$. Furthermore,
for either choice (``$K\pm\pi$'') 
the cross section for $pp\rightarrow p\Sigma^+K^0$ is 
about 3.3 times larger then the one for $pp\rightarrow p\Sigma^0K^+$. 
Note that the experimental evidence at higher energies suggests a ratio
of around 1 for those channels. 

\begin{figure}[htb]
\vskip 5.0cm
\includegraphics{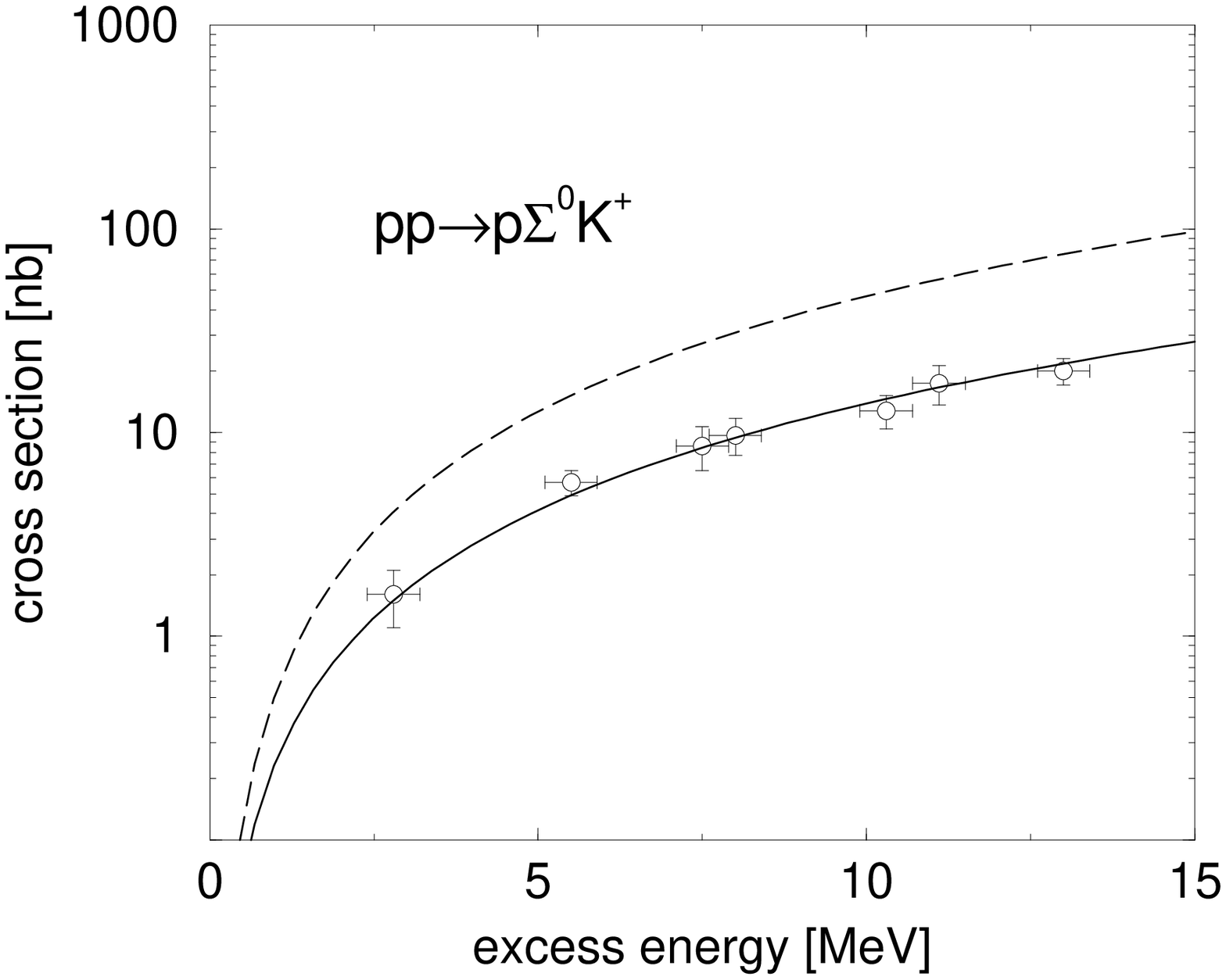}
\includegraphics{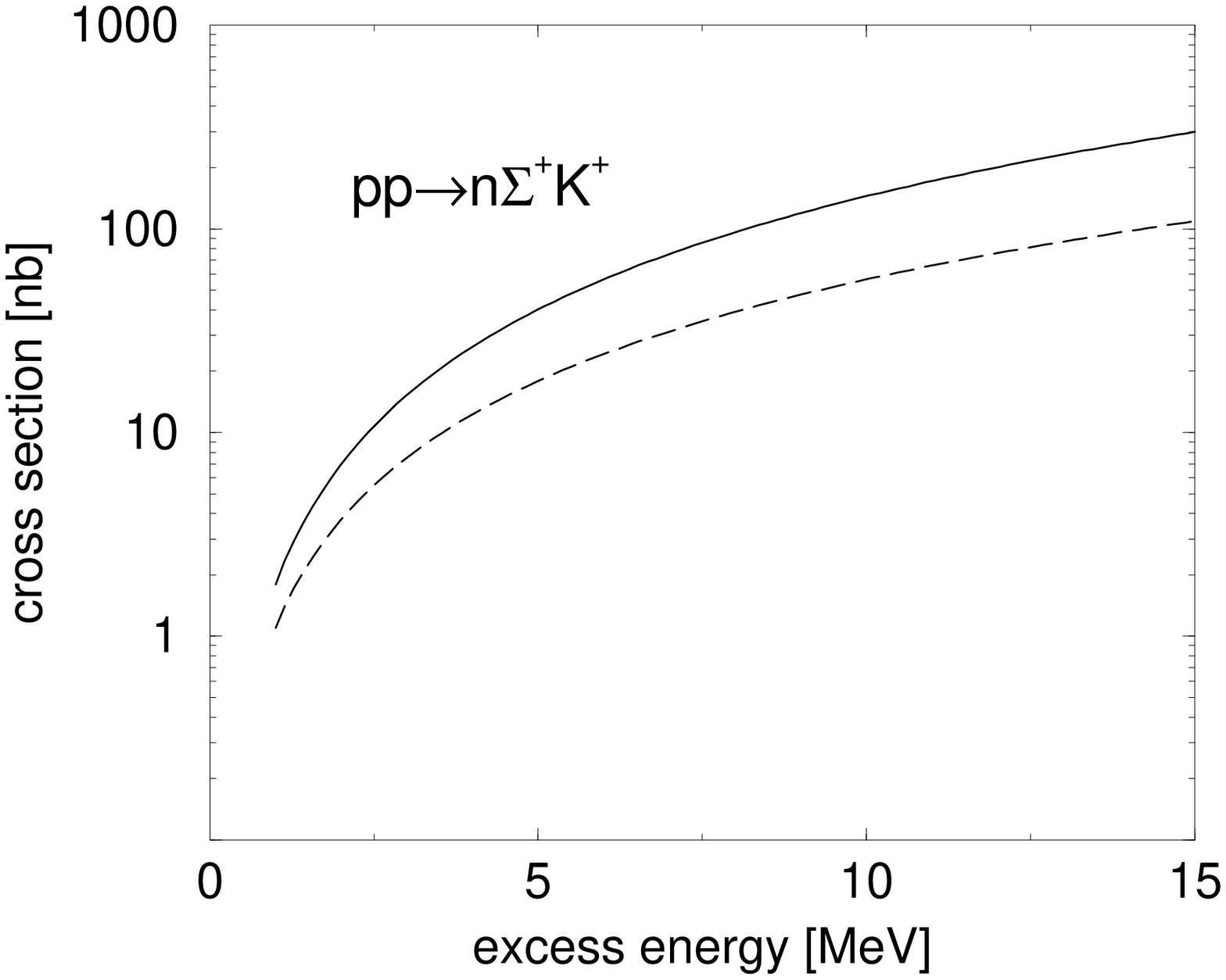}
\caption{Total cross sections for the reactions
$pp\rightarrow p\Sigma^0K^+$ and $pp\rightarrow n\Sigma^+K^+$. 
The solid curve corresponds to the choice ``$K-\pi$'' and the dashed curve
to ``$K+\pi$'', cf. text. All curves are normalized by a factor of 0.3.
The experimental data are from Ref.~\protect\cite{Sew}. }
\label{fig:on-shell}
\end{figure}

\def\Nucl{Nucl.\ }
\def\Phys{Phys.\ }
\def\Rev{Rev.\ }
\def\Lett{Lett.\ }
\def\PL{\Phys\Lett}
\def\PLB{\Phys\Lett B}
\def\NP{\Nucl\Phys}
\def\NPA{\Nucl\Phys A}
\def\NPB{\Nucl\Phys B}
\def\NPBS{\Nucl\Phys (Proc.\ Suppl.\ )B}
\def\PR{\Phys\Rev}
\def\PRL{\Phys\Rev\Lett}
\def\PRC{\Phys\Rev C}
\def\PRD{\Phys\Rev D}
\def\RMP{\Rev  Mod.\ \Phys}
\def\ZP{Z.\ \Phys}
\def\ZPA{Z.\ \Phys A}
\def\ZPC{Z.\ \Phys C}
\def\AOP{Ann.\ \Phys}
\def\PRep{\Phys Rep.\ }
\def\ANP{Adv.\ in \Nucl\Phys Vol.\ }
\def\PTP{Prog.\ Theor.\ \Phys}
\def\PTPS{Prog.\ Theor.\ \Phys Suppl.\ }
\def\PL{\Phys \Lett}
\def\JPF{J.\ Physique}
\def\FBSS{Few--Body Systems, Suppl.\ }
\def\IJMP{Int.\ J.\ Mod.\ \Phys A}
\def\NuCi{Nuovo Cimento~}

\end{document}